\documentclass[10pt]{article}

\usepackage{amsmath}
\usepackage{amssymb}
\usepackage{graphics}
\usepackage{rotating}
\usepackage{cite}
\usepackage{color}
\usepackage{fancybox}

\def\0{\mbox{\tiny $0$}}
\def\1{\mbox{\tiny $1$}}
\def\2{\mbox{\tiny $2$}}
\def\3{\mbox{\tiny $3$}}
\def\4{\mbox{\tiny $4$}}
\def\5{\mbox{\tiny $5$}}
\def\6{\mbox{\tiny $6$}}
\def\7{\mbox{\tiny $7$}}
\def\8{\mbox{\tiny $8$}}
\def\9{\mbox{\tiny $9$}}

\def\T{\mbox{\tiny $T$}}

\def\EE{\mbox{\tiny EE}}
\def\TB{\mbox{\tiny TB}}

\def\c{\mbox{\tiny $c$}}

\title{\shadowbox{\large \bf RESONANT LASER TUNNELLING}}

\author{
\small  Stefano De Leo\thanks{Department of Applied Mathematics,
State University of Campinas, Brazil [deleo@ime.unicamp.br] } \,\,
and\, Pietro Rotelli\thanks{Department of Physics, University of
Salento and INFN Lecce, Italy [rotelli@le.infn.it]}}

\date{\small
\fcolorbox{black}{yellow} {\color{red} $\bullet$ {\color{black}{
{\footnotesize  {\sc European Physical Journal D} {\bf 65}, 563-570 (2011)}}}
{\color{red}{$\bullet$}} } }

\begin{document}
%

\maketitle

\vspace*{-.7cm}

\begin{abstract}
\noindent We propose an experiment involving a gaussian laser
tunneling through a twin barrier dielectric structure. Of
particular interest are the conditions upon the incident angle for
resonance to occur. We provide some numerical calculations for a
particular choice of laser wave length and dielectric refractive
index which confirm our expectations.
\end{abstract}












\section*{\normalsize I. INTRODUCTION}

In  previous articles\cite{del1,del2}, we  pointed out the
analogies between laser interactions with a dielectric block and
non-relativistic quantum mechanics (NRQM). Tests of the latter
physics should be easier through use of this optical analog, if
only because we do not have to deal with probability amplitudes.
One example is the transition in NRQM barrier diffusion from
complete coherence to complete incoherence. Resonance phenomena
occurs for the former limit but is absent in the latter. For
example, the condition for coherence in {\em one-dimensional} NRQM
diffusion is that the spatial size of the single incoming wave
packet be much larger than the barrier size\cite{del3,del4}. With
coherence a single wave packet is transmitted. With complete
incoherence, unlimited numbers of wave packets are transmitted at
regular intervals. Whence, incoherence phenomena are absent when
considering only a \emph{single} incoming plane wave because of
its infinite size. Within the laser/dielectric analogy the wave
packet size is substituted by the laser beam transverse spreads.
Interacting with a dielectric block, as expected from geometric
optics, multiple reflections occur\cite{book1}. Coherence attains
only if overlaps dominate in the outgoing laser spots. One
important difference between optics and NRQM is that the optical
situation is {\em stationary}. Time in NRQM is essentially
replaced by a spatial variable in optics. This offers some
significant advantages, e.g. even low intensity beams can be
detected by long time exposures in detectors.

The fact that optical experiments provide an easy way to test quantum mechanical
 predictions,  since the early  1970's,  stimulated experimental and theoretical studies  on quantum-optical analogies\cite{r1,r2,r3,r4,r5,r6,r7,r8,r9}. For clear and detailed  reviews on the
correspondence between optics and quantum mechanics, we refer the reader to the works of Dragoman\cite{dra} and Longhi\cite{lon}.

Since any realistic setup is three dimensional, we must also work
theoretically in three dimensions. Even when all interesting
physics is dominantly bi-dimensional, e.g. the $y$-$z$ plane
defined by the incoming central laser direction (chosen by us as the $z$
direction) and the normal to a structured dielectric system.
There is a benefit in the extension beyond a simple one
dimensional study to more dimensions. It means that the variation
in ``energy'' in one-dimensional NRQM\cite{book2} can now be
achieved by merely changing the incident angle. Indeed, it is the
normal momentum component which now plays the role of energy in one dimension. For example, in
optics,  we can, under certain conditions, transit from diffusion to tunneling by simply
rotating the dielectric  structure relative to the incoming beam.

A dielectric-air-dielectric system mimics a potential
barrier. Experiments realized by a double prism
arrangement\cite{dp1,dp2} have recently discussed the phenomenon
of frustrated total internal reflection. In this paper, we go beyond a
dielectric-air-dielectric system and consider
a double prism plus a narrow slab separated by small air gaps of
\emph{equal size} (see Fig.\,1). This reproduces a twin barrier
situation\cite{esp,rec,del5} which should exhibit one of the most
intriguing results of NRQM, the resonant tunneling
effect\cite{rpt1,rpt2}. Notwithstanding the expected attenuation
of each barrier (air gap) upon the laser beam, for certain angular
incidences the transition probability through twin barriers
becomes unity, i.e. \emph{total transmission} occurs. We determine
the conditions for this phenomenon and reproduce sample results
from a numerical calculation which necessarily takes into account
the finite angular spread of any laser beam. By eliminating the
intermediate slab, we reproduce the result of tunnelling through a
single barrier\cite{tun1,tun2}. The twin-barrier transmission
results are attenuated by the entry/exit transmission amplitudes.
All three must be folded into the laser momentum amplitude before
integrating. This yields the outgoing laser beam amplitude and
consequent intensity. For our analysis it is advisable to avoid
the complication of multiple outgoing beams produced by
incoherences within the dielectric prisms. This is probably best
achieved by considering only small incident angles $\alpha$. In
principle it could also be achieved by reducing the depth of the
prisms. This would facilitate the search for resonance at larger
incident angles. Another effect that would also help in the use of larger
incident angles, without lose of coherence within the prisms, is
the fact that the $y$-profile of the laser beam is increased with
$\alpha$ as 1/cos$\alpha$. The $x$-profile is constant throughout
and plays no role in the question of coherence. In the next
section, we describe our theoretical treatment of the gaussian
laser. In the subsequent section, we describe our proposed twin
barriers structure. In section IV, we provide the
transmission/reflection probabilities for entry, exit and for the
twin barriers. Section V contains our numerical transmission
results for a chosen laser wave length and dielectric refractive
index. The results confirm our phenomenological expectations. Our
conclusions are drawn in the final section.

\section*{\normalsize II. THE GAUSSIAN LASER}

 Discussion of the laser is simplified by first
concentrating upon the electric amplitude it produces and,
wherever possible treating the amplitude as if it where a scalar.  This amplitude is
a convolution in wave number of plane waves. For
different plane wave directions the polarization vector
necessarily changes since it lies in the plane perpendicular to
the wave direction $\boldsymbol{k}$. However, for a \emph{sharp}
gaussian in $k_x$ and $k_y$ we may assume to a good approximation
that this vector (if linearly polarized) points in a fixed
direction,
\[\boldsymbol{E}(x,y,z)=\left\{\,E_s(x,y,z)\,,\,E_p(x,y,z)\,\right\}\,\,,\]
 where $E_s$ is the component of the electric field
perpendicular to the plane of incidence ($s$-polarization) and $E_p$ is the component
perpendicular to this plane ($p$-polarization).

In this paper, we shall present the discussion of resonant tunneling for a gaussian He-Ne laser, $\lambda=633$ nm, of width $\mbox{w}_{\0}(=2$mm)  which, in free space, propagates with principal axis along the
$z$-direction, i.e.
\begin{equation}
\label{gaus}
 G(k_x,k_y)=\exp\left[\,-\,\frac{\mbox{w}_{\0}^{\2}}{4}\,
\left(\,k_x^{^{2}}+k_{y}^{^{2}}\, \right)\right]\,\,.
\end{equation}
To simplify our presentation, we shall suppose that the laser beam passes
through a linear polarizer which selects the $s$-polarization. This allows, as shown in the next section, to directly use the QM results. The electric field can be thus treated as  a scalar,
\[\boldsymbol{E}(x,y,z)=\left\{\,E_s(x,y,z)\,,\,E_p(x,y,z)\,\right\}=\left\{\,E(x,y,z)\,,\,0\,\right\}
\,\,,\]
where

\begin{equation}
\label{las0}
 E(x,y,z)  =
E_{\0}\,\frac{\mbox{w}^{\2}_{\0}}{4\,\pi}\,\, \int
\mbox{d}k_x\,\mbox{d}k_y\,\,G(k_x,k_y)\,e^{i\,(k_x\,x+k_{y}\,y+k_z\,z)}\,\,.
\end{equation}
Observing that $k\mbox{w}_{\0}=2\pi\mbox{w}_{\0}/\lambda \gg 1$, we find the following analytic analytic expression for
the free space intensity,
\begin{equation}
\mbox{I}(x,y,z) = \mbox{I}_{\0}\,
\frac{ \mbox{w}_{\0}^{\2}}{\mbox{w}^{\2}(z)}\,\exp\left[\,-\,2\,\,
\frac{x^{^{2}}+y^{^{2}}}{\mbox{w}^{\2}(z)} \right]\,\,\,,\,\,\,\,\,\,\,\,\,\,
\mbox{w}(z)=\mbox{w}_{\0}\,\sqrt{1+\left(\,\frac{\lambda\,\,z}{\pi\,\mbox{w}_{\0}^{\2}}\,
\right)^{^{2}}}\,\,.
\end{equation}

\section*{\normalsize III. LAYOUT OF TWIN BARRIER APPARATUS}

In Fig.\,2, we show the $y$-$z$ overview of a prism which forms a part
of a twin barrier structure. The incident ray corresponding to the
central direction of the laser travels along the $z$-axis. The
incident angle with the normal, $z_*$, of the leading face of the
dielectric is  denoted by $\alpha$. The internal direction $\beta$ satisfies Snell's law,
\begin{equation}
\sin\alpha=n\,\sin\beta\,\,.
\end{equation}
Before to derive the conditions for total internal reflection (TIR), let us establish
the connection between optical and quantum mechanical problems. This connection
will be the starting point for the calculation of the transmission probabilities presented in the next section.

The Fresnel formulas for the reflected and transmitted waves obtained when s-polarized light  moves from air to a dielectric of a given refractive index $n$ are\cite{book1}
\begin{equation}
R_s=\frac{\cos \alpha - n \cos \beta}{\cos \alpha + n \cos \beta}\,\,\,\,\,\,\,\,\mbox{and}\,\,
\,\,\,\,\, T_s=\frac{2 \, \cos \beta}{\cos \alpha + n \cos \beta}\,\,.
\end{equation}
The air dielectric system is, for s-polarized incoming waves, the NRQM analog of a
step potential. Indeed, for a step potential with
a discontinuity perpendicular to the $z_*$-axis we obtain the following reflection and transmission coefficients\cite{book2}
 \begin{equation}
R_{step}=\frac{k_{z_*} - q_{z_*} }{k_{z_*} - q_{z_*}}\,\,\,\,\,\,\,\,\mbox{and}\,\,
\,\,\,\,\, T_{step}=\frac{2 \, k_{z_*}}{k_{z_*} - q_{z_*}}\,\,.
\end{equation}
Now, observing that for plane wave
\[k_{z_*}=k\cos \alpha\,\,\,\,\,\,\,\mbox{and}\,\,\,\,\, q_{z_*}=\sqrt{n^{^2}k^{^2}-k_{y_*}}=\sqrt{n^{^2}k^{^2}-k^{^2}\sin^{\2}\alpha}=n\,k\,\cos\beta\,\,,\]
we find $R_{step}=R_s$ and $T_{step}=T_s$. This analogy can be
claimed even if there are subsequent structures for we now know
that all phenomena can be calculated as successive events even
when coherence reigns. For the above situation there are always
reflected and transmitted beams since
the momentum vector is greater within the dielectric than in air. We are in the analogy of
QM {\em diffusion}.

Passing from the dielectric to air, we have a situation where for angles
greater than a critical incident angle $\gamma_c$,
\begin{equation}
n\,\sin \gamma_c =1\,\,,
\end{equation}
total internal reflection
(TIR) occurs because Snell's law cannot be satisfied. If the air
region is of finite dimensions, we are in the analogy of QM
{\em tunneling}.

The twin barrier structure, see Fig.\,1, is created by air zones
separating two prisms from a narrow slab made of the same material
as the prisms. Since the resonance phenomena that we seek to
reproduce requires coherence, the total distance between the two
prisms must be much smaller than the laser beam size (with one notable exception - head
on incidence). For certain
angles, as we shall demonstrate in the next section, transmission
through the twin barriers will be unity (resonance). However, if the barriers
are not of the same size resonance is attenuated. The direction
normal to this twin barrier structure we call $\tilde{z}$. The
overall transmission (reflection) amplitudes comprising of the
entry step, twin barrier and exit step must multiply the gaussian
modulation before integrating over wave number plain waves. In
general this can only be done numerically. Obviously, the entry
and exit step amplitudes will produce a reduction in laser
intensity even at resonance. This should not be a serious problem
for a steady state intensity, which can be measured over an
arbitrary  time interval. As a check of our results, the simpler
single barrier amplitude can be derived from the twin barrier one
by setting the intermediate-slab thickness to zero. The resulting
amplitude is that for a barrier width equal to the \emph{sum} of
our twin barrier sizes.\\

Before we calculate the transmission amplitudes, it is useful to derive the conditions for TIR.  To do so we must consider
two distinct cases.\\

\noindent $\bullet$
\textbf{Case (a)}. Shown in Fig\,2a.  From the internal triangle with angles
$\gamma$, $\pi/2 -\beta$, and $\pi/4$, we find
\begin{equation}
\gamma = \pi/4+\beta\,\,.
\end{equation}
TIR occurs for values $\gamma >\gamma_c$. This implies the following constraint on $\beta$
\begin{equation}
\label{ca1}
\sin 2\,\beta > \sin \left(\,2\,\gamma_c - \pi/2\,\right)\,\,.
\end{equation}
Observing that $\sin 2\,\beta>0$, the previous condition is always satisfied when $\gamma_c<\pi/4$  since the right hand side is negative. Thus for $n>\sqrt{2}$ only TIR occurs. Let us now investigate the constraint on $\alpha$ for TIR in the case $n<\sqrt{2}$. From Eq.(\ref{ca1}), we get
\begin{equation}
\label{ca2}
\sin 2\,\beta > 2\,\sin^{\2} \gamma_c - 1 = \frac{2-n^{\2}}{n^{\2}}\,\,\,\,\,\Rightarrow\,\,\,\,\,
\cos 2\,\beta < \frac{2\,\sqrt{n^{\2}-1}}{n^{\2}}\,\,\,\,\,\Rightarrow\,\,\,\,\,\sin^{\2}\alpha >\frac{n^{\2}- 2\,\sqrt{n^{\2}-1}}{2}\,\,.
\end{equation}
Summarizing, in  case (a), the conditions on the incident angle
$\alpha$ which guarantee TIR are
\begin{equation}
\begin{array}{lcc}
n>\sqrt{2} & : &  \forall \,\alpha\,\,,\\
n<\sqrt{2} & : &  \alpha > \arcsin \sqrt{\displaystyle{\frac{n^{\2}- 2\,\sqrt{n^{\2}-1}}{2}}}\,\,.
\end{array}
\end{equation}
\noindent $\bullet$
\textbf{Case (b)}. Shown in Fig\,2b.  This is obtainable by flipping
the prism (and consequently the full dielectric set up) with
respect to the incoming central laser direction. Alternatively,
and perhaps more practically, it is obtained from the previous
case by flipping the laser beam about the normal to the prism.
From the internal triangle with angles
$\pi/2-\theta$, $\pi/2 -\beta$, and $\pi/4$, we find
\begin{equation}
\gamma = \pi/4-\beta\,\,.
\end{equation}
TIR ($\gamma >\gamma_c$) imposes the following constraint on $\beta$
\begin{equation}
\label{cb1}
\sin 2\,\beta < \sin \left(\,\pi/2-2\,\gamma_c\,\right)\,\,.
\end{equation}
Observing that $\sin 2\,\beta>0$, the previous condition is never satisfied if $\gamma_c>\pi/4$  since the right hand side is negative. Thus for $n<\sqrt{2}$ only diffusion occurs. For $n>\sqrt{2}$, from Eq.(\ref{cb1}), we have
\begin{equation}
\label{cb2}
\sin 2\,\beta < 1- 2\,\sin^{\2} \theta_c  = \frac{n^{\2}-2}{n^{\2}}\,\,\,\,\,\Rightarrow\,\,\,\,\,
\cos 2\,\beta > \frac{2\,\sqrt{n^{\2}-1}}{n^{\2}}\,\,\,\,\,\Rightarrow\,\,\,\,\,\sin^{\2}\alpha <\frac{n^{\2}- 2\,\sqrt{n^{\2}-1}}{2}\,\,.
\end{equation}
Finally, in case (b), the conditions on the incident angle
$\alpha$ which guarantee TIR are
\begin{equation}
\begin{array}{lcc}
n<\sqrt{2} & : &  \emptyset\,\,,\\
n>\sqrt{2} & : &  \alpha < \arcsin \sqrt{\displaystyle{\frac{n^{\2}- 2\,\sqrt{n^{\2}-1}}{2}}}\,\,.
\end{array}
\end{equation}

\section*{\normalsize IV. TRANSMISSION  PROBABILITIES}

The principal transmission  amplitude emerging from the far edge of
the dielectric structure as sketched in Fig.\,1, is a product of
three separate transition amplitudes, the entry and exit
transitions and the twin barrier transition (assumed coherent in
this study). Reflected amplitudes also occur at each stage and
will in general give rise to secondary and higher transition
amplitudes which we ignore for simplicity. Since the perpendicular
momentum or wave number is of such relevance much of what follows
involves the use of appropriate rotation matrices. We start with
the entry step and consider, prior to convolution, an incoming
plane wave.\\

\noindent $\bullet$ {\bf Entry [air/dielectric interface].}\\

\noindent At this interface the standard one-dimensional step
results hold with the relevant momentum given by the $z_*$
component of $\boldsymbol{k}$. The $z_*$ direction is normal to
the interface and in terms of our original axis (chosen so that
the principal direction coincides with the $z$-axis) it
corresponds to a rotation of $\alpha$ around the $x$-axis,
\[
\left(\begin{array}{c} z_*\\ y_* \end{array} \right) = \left(
\begin{array}{rr} \cos \alpha & -\,\sin \alpha \\
\sin\alpha & \cos \alpha
\end{array}   \right)\left(\begin{array}{c} z\\ y \end{array}
\right)\,\,.
\]
The wave number components of the incoming laser beam which
propagates in air can be expressed in the new axis in terms of the
wave number $\boldsymbol{k}$,
\begin{equation}
\begin{array}{lcl}
k_{x_*} & = & k_x\,\,,   \\
k_{y_*} & = & k_z \sin \alpha + k_y \cos \alpha = k\,\sin \alpha +
k_y \cos \alpha +
\mbox{O}[k_x^{\2},k_y^{\2}]\,\,,\\
 k_{z_*} & = & k_z\cos \alpha - k_y \sin \alpha=k\,\cos \alpha - k_y \sin \alpha +
\mbox{O}[k_x^{\2},k_y^{\2}]\,\,.
\end{array}
\end{equation}
The wave number components of the laser beam which propagates in
the dielectric are
\begin{equation}
\begin{array}{lcl}
 q_{x_*} & = & k_{x_*}\,\,,  \\
q_{y_*} & = & k_{y_*}\,\,, \\
q_{z_*} & = & \sqrt{n^{\2}k^{\2}-k_{x_*}^{\2}-k_{y_*}^{\2}}=
n\,k\,\cos\beta - k_y\,\tan\beta \cos\alpha+
\mbox{O}[k_x^{\2},k_y^{\2}] \,\,.
\end{array}
\end{equation}
Now the expressions for the transmission (reflection) amplitudes
have phases depend upon the value of $z_*$ for the step. Our laser
formulas are written with the origin as the central point of
minimum transverse size of the laser beam. We denote the $z_*$
distance of the interface from this by $a_*$ (see
Fig.\,1). With these definitions, we recall the step results
\begin{equation}
R^{^{(1,n)}}_{a_*}=\frac{k_{z_{*}}-\,q_{z_*}}{k_{z_*}+\,
q_{z_*}}\,\,\,e^{2\,i\,k_{z_*}a_*}\,\,\,\,\,\,\mbox{and}
\,\,\,\,\,\,\,T^{^{(1,n)}}_{a_*}=\frac{2\,k_{z_{*}}}{k_{z_*}+\,
q_{z_*}}\,\,\,e^{i\,(k_{z_*}-\,q_{z_*})a_*}\,\,.
\end{equation}

\noindent $\bullet$ {\bf Exit [dielectric/air
interface].}\\

\noindent For the exiting stage dielectric/air interface at $z_*=b_*$, we
need only interchange in the entry results $k_{z_*}$ with
$q_{z_*}$ and change $a_{*}$ with $b_*$,
\begin{equation}
R^{^{(n,1)}}_{b_*}=\frac{q_{z_{*}}-\,k_{z_*}}{q_{z_*}+\,
k_{z_*}}\,\,\,e^{2\,i\,q_{z_*}b_*}\,\,\,\,\,\,\mbox{and}
\,\,\,\,\,\,\,T^{^{(n,1)}}_{b_*}=\frac{2\,q_{z_{*}}}{q_{z_*}+\,
k_{z_*}}\,\,\,e^{i\,(q_{z_*}-\,k_{z_*})b_*}\,\,.
\end{equation}
For our calculation of the transmission amplitude, we shall need the
product of the entry and exit transmission amplitudes, $T_{\EE}$,
\begin{equation}
T_{\EE}(k_x,k_y) = T^{^{(1,n)}}_{a_*} T^{^{(n,1)}}_{b_*} = \frac{4\,k_{z_{*}}q_{z_{*}}}{\left(\,k_{z_*}+\,
q_{z_*}\right)^{^{2}}}\,\,\,e^{i\,(q_{z_*}-\,k_{z_*})D_*}\,\,,
\end{equation}
where $D_*=b_*-a_*$. In Fig.\,3, we show the values of the above product for
different values of refractive index $n$ as a function of the
incident angle $\alpha$. The values of $k_x$ and $k_y$ have been
chosen as zero here, so these curves strictly apply to the central
laser ray. The dependence is mild (particularly for small incident
angles) compared, for example, to that induced by resonance (see
below). Its main effect will be that of reducing the emerging
laser beam intensity compared to the ingoing laser beam, even at
the peak of resonance when the twin barrier amplitude is unity. As
pointed out in the previous section, this should not constitute a
serious problem.\\

\noindent $\bullet$ {\bf Central structure [dielectric/air/dielectric/air/dielectric
interface].}\\

\noindent
 For resonance phenomena in tunnelling, we consider the
central structure as small or comparable to the laser dimensions
and give the overall transmission (reflection) amplitudes for
this. Before doing so it is convenient to rotate the axis anew
about $x$ to define the $\tilde{z}$ direction normal to the
central structure
\[
\left(\begin{array}{c} \tilde{z}\\ \tilde{y} \end{array} \right) =
\frac{1}{\sqrt{2}}\, \left(
\begin{array}{rr} 1 & -\,1 \\
1 & 1
\end{array}   \right)\,\left(\begin{array}{c} z_*\\ y_*
\end{array} \right)\,\,.
\]
In terms of the new axis,  the dielectric wave numbers become
\begin{equation}
\begin{array}{lcl}
q_{\tilde{x}} & = & q_{x_*}\,\,,     \\
q_{\tilde{y}} & = &
(q_{z_*}+\,q_{y_*})/\sqrt{2}=n\,k\,\sin\gamma+k_y\,\cos\gamma\,
\cos\alpha\,/\cos\beta +
\mbox{O}[k_x^{\2},k_y^{\2}] \,\,,
 \\
 q_{\tilde{z}} & = & (q_{z_*}-\,q_{y_*})/\sqrt{2}=n\,k\,\cos\gamma-k_y\,\sin\gamma\,
\cos\alpha\,/\cos\beta +
\mbox{O}[k_x^{\2},k_y^{\2}]\,\,,
\end{array}
\end{equation}
and the air wave numbers
\begin{equation}
\begin{array}{lcl}
k_{\tilde{x}} & =
 &q_{\tilde{x}}\,\,.  \\
k_{\tilde{y}} & = & q_{\tilde{y}}\,\,, \\
 k_{\tilde{z}} & = & \sqrt{k^{\2}-q_{\tilde{x}}^{\2}-q_{\tilde{y}}^{\2}}
= k\,\sqrt{1-n^{\2}\sin^{\2}\gamma} -
n\,k_y\,\displaystyle{\frac{\sin \gamma \cos
\gamma}{\sqrt{1-n^{\2}\sin^{\2}\gamma}}
\frac{\cos\alpha}{\cos\beta}} + \mbox{O}[k_x^{\2},k_y^{\2}]\,\,.
\end{array}
\end{equation}
For  $n\, \sin \gamma>1$, we have tunneling. In this case, we can
rewrite $k_{\tilde{z}}=i\,|k_{\tilde{z}}|$ where
\[
|k_{\tilde{z}}| = k\,\sqrt{n^{\2}\sin^{\2}\gamma-1} +
n\,k_y\,\displaystyle{\frac{\sin \gamma \cos
\gamma}{\sqrt{n^{\2}\sin^{\2}\gamma-1}}
\frac{\cos\alpha}{\cos\beta}} + \mbox{O}[k_x^{\2},k_y^{\2}]
\]
The single barrier reflection and transmission coefficients are\cite{book2}
\begin{equation}
R  = -\,i\,
\frac{q_{\tilde{z}}^{\2}+|k_{\tilde{z}}|^{^{2}}}{2\,q_{\tilde{z}}\,|k_{\tilde{z}}|}\,\,
\sinh\left(|k_{\tilde{z}}|\,\widetilde{L}\right)\, e^{i\,\varphi}
\,/\,\mathcal{F}\,\,\,\,\,\,\,\,\,\mbox{and}\,\,\,\,\,\,\,\,\,
T  =  e^{i\,(\varphi - q_{\tilde{z}}\,\widetilde{L})}\,/\,\mathcal{F}\,\,,
\end{equation}
where
\[
\mathcal{F} =  \sqrt{1+\left[
\frac{q_{\tilde{z}}^{\2}+|k_{\tilde{z}}|^{^{2}}}{2\,q_{\tilde{z}}\,|k_{\tilde{z}}|}\,
\sinh\left(|k_{\tilde{z}}|\,\widetilde{L}\right)\right]^{^{2}}}\,\,\,,\,\,\,\,\,\,\,\,\,
\varphi =  \arctan\left[\,
\frac{q_{\tilde{z}}^{\2}-|k_{\tilde{z}}|^{^{2}}}{2\,q_{\tilde{z}}\,|k_{\tilde{z}}|}\,\,
\tanh\left(|k_{\tilde{z}}|\,\widetilde{L}\right)\,\right]\,\,,
\]
and $\widetilde{L}$ is the dimension along the $\tilde{z}$ axis of the air gap (see the figure below). The amplitude due to a  back and forth reflection within the dielectric slab region produces the following loop factor
\[
R\,\exp[\,2\,i\,q_{\tilde{z}}\,\tilde{b}\,]\,\times\,
R\,\exp[\,-\,2\,i\,q_{\tilde{z}}\,\tilde{a}\,] =
R^{^{2}}\,\exp[\,2\,i\,q_{\tilde{z}}(\tilde{b}\,-\,\tilde{a})\,]=-\,\left|
R\right|^{^{2}}\,\exp[\,2\,i\,(q_{\tilde{z}}\,\widetilde{D}\,+\,\varphi)\,]\,\,,
\]
where $\widetilde{D}$ is the dimension along the  $\tilde{z}$ axis of the
dielectric slab (see amplification in Fig.\,1). Thus performing the sum over successive loop
contributions to the twin barrier amplitude, $T_{\TB}$, yields
\begin{equation}
T_{\TB}(k_x,k_y)=\frac{T^{^{2}}}{1+\left|
R\right|^{^{2}}\,\exp[\,2\,i\,(q_{\tilde{z}}\,\widetilde{D}\,+\,\varphi)\,]}\,\,.
\end{equation}
The above  result can also be calculated through a
standard matrix approach. A check of this result can be made by
setting $\widetilde{D}=0$. This yields the single barrier result for a barrier
{\em size} of $2\,\widetilde{L}$ to be compared with that for the size $\widetilde{L}$
barrier results used in the above derivation. The results are in
accord.

The resonance conditions are achieved when the phase factor
multiplying $|R|^{^{2}}$ equals $-1$. The denominator is then equal
to $|T|^{^{2}}$ and the twin barrier amplitude reduces to a mere
phase and hence of unit modulus. In general the denominator will
far exceed the numerator, which for sufficiently large $\widetilde{L}$ is
proportional to a decreasing exponential in $\widetilde{L}$. In Fig.\,4, we show
some explicit examples of this phenomena. In this figure, we have set arbitrarily $k_x=0$ and
$k_y=0$ and plotted $|T_{\TB}|$ versus the incident angle $\alpha$ for a refractive index $n=\sqrt{3}$. Note that this angle is not the impinging angle upon the twin
barrier structure itself (see Fig.\,2). The three curves are for
$\{k\widetilde{L},k\widetilde{D}\}=\{1.5,100\},\,\{1,100\},\,\{1,500\}$. In each the resonance structure is clearly manifest. The peaks are very close to each other in the
latter case. The first two plots demonstrate that for a fixed $\widetilde{D}$ the resonance peaks are narrower as $\widetilde{L}$ increases. As an aside, we recall that resonance requires the existence of a non null $\widetilde{D}$. The resonance condition cannot be obtained by $\varphi$ alone. For tunnel resonance one needs multiple barriers.

\section*{\normalsize V. NUMERICAL ANALYSIS}

Our interest in this paper is the outgoing transmission amplitude
from the twin barrier dielectric set-up described in the previous
section and sketched in Fig.1. In particular, we are interested in
the resonance phenomena displayed within the tunnelling regime,
i.e. for TIR.   TIR introduced for a step potential is somewhat of a
misnomer for finite barriers. Indeed resonance, when it occurs, is
the extreme counterexample since it represents {\em total transmission} not
total reflection.  The electric field for  the transmitted laser is obtained by
multiplying the gaussian $G(k_x,k_y)$ in Eq.(\ref{las0}) by
$T_{\EE}(k_x,k_y)$ and $T_{\TB}(k_x,k_y)$,
\[
 E_{\T}(x,y,z)  \approx
E_{\0}\,\frac{\mbox{w}^{\2}_{\0}}{4\,\pi}\,\, \int
\mbox{d}k_x\,\mbox{d}k_y\,\,G(k_x,k_y)\,T_{\EE}(k_x,k_y)\,T_{\TB}(k_x,k_y)\,
e^{i\,(k_x\,x+k_{y}\,y +k_z\,z)}\,\,.
\]
Observing that  $k\mbox{w}_{\0}\gg1$ and that the dependence on $k_x$ in $T_{\EE}\,T_{\TB}$ is only of second order, whereas the $k_y$ dependence is of the first order, we can simplify the transmitted electric field as follows
\begin{equation}
\label{lasT2}
 \left|E_{\T}(x,y,z)/E_{\0}\right|   \approx
\frac{\mbox{w}^{\2}_{\0}}{2\,\sqrt{\pi}\,\mbox{w}(z)}\,\,\exp[-\,x^{\2}/\mbox{w}^{\2}(z)] \left|\,\int
\mbox{d}k_y\,\,\,G_{\T}(k_y)\,
\exp\left[\,i\,\left(k_{y}\,y-
\frac{k_y^{^{2}}}{2k}\,z\,\right)\, \right]\,\right|\,\,,
\end{equation}
where
\[G_{\T}(k_y)=\exp[-k_y^{^{2}}\mbox{w}_{\0}^{\2}/4]\,\,T_{\EE}(0,k_y)\,T_{\TB}(0,k_y)\,\,.\]
In Fig.\,5, we plot $\left|G_{\T}(k_y)\right|$ as a function of $k_y$ around the angular resonance for the following choices  $\{k\widetilde{L},k\widetilde{D}\}=\{3,10^{^{2}}\},\,\{1,10^{^{2}}\},\,\{1,10^{^{4}}\},\, \{1,5\cdot 10^{^{4}}\}$. Also plotted in the same figure is the incoming gaussian with $k_x=0$, i.e. $G(0,k_y)$. The effect of $T_{\EE}(0,k_y)\,T_{\TB}(0,k_y)$ reduces the peak values of each curve and for $\{k\widetilde{L},k\widetilde{D}\}=\{3,10^{^{2}}\}$ shifts the peak significantly to a positive value of $k_y$. In the case with $\{k\widetilde{L},k\widetilde{D}\}=\{1,5\cdot 10^{^{4}}\}$, we see the presence of multiple peaks. Observe that the values of $\widetilde{L}$ are  very small if compared to the laser beam size, $k\widetilde{L}=1$ corresponds to $\widetilde{L}=0.1\,\mu\mbox{m}$.

The spatial size of the laser is set by $\mbox{w}_0$ and is of $2$ mm. For
distances between laser source, dielectric structure and
measurement apparatus  of the order of a few meters we can readily
neglect spatial spreading due to the gaussian, although this does
not necessarily apply (as we shall see) to the effect of twin
barrier resonance upon the beam. As we have argued in depth
elsewhere, resonance phenomena requires a twin barrier structure
smaller in size than the beam size. This means that $\widetilde{D}+2\widetilde{L}\ll 2 \mbox{mm}$.
Since for other reasons $\widetilde{L}$
must be much smaller (of the order of $\lambda$) this limits $\widetilde{D}$
to less than a mm. For greater values of $\widetilde{D}$ the resonance peaks proliferate so that the
gaussian laser would encompass multiple resonance peaks. In this
case varying the incident angle produces little or no consequences
and the resonance effect is averaged out.

With the same selection of values of $\{\widetilde{L},\widetilde{D}\}$ used in Fig.\,5, we have performed a numerical calculation of $\left|E_{\T}(x,y,z)\right|^{\2}$. In Fig.\,6, we plot the exit contour of the laser intensity  in the $x$-$y$ plane at the $1/e^{\2}$ width.  The results are, of course, correlated to those of Fig.\,5. For example, a narrow curve in $k_y$ such as in Fig.\,5a produces a large spread in the $y$ contour in Fig.\,6a. The $x$-spread  is the same in all of the  Fig.\,6 because the $x$-integral in Eq.\,(\ref{lasT2})  factorizes (after ignoring second order corrections in $k_x$). In Fig.\,6d, we display for comparison the incident laser beam with peak intensity I$_{\0}$. The valures of the peak intensities (all close to $x=y=0$) are also given in each figure. Fig.\,5b corresponding to $\{k\widetilde{L},k\widetilde{D}\}=\{1,10^{^{2}}\}$ is notable because it almost reproduces the incident laser reduced in intensity by a factor of $1/2$. We do not display the more complex case corresponding to Fig.\,5d with its multiple peaks.

\section*{\normalsize VI. CONCLUSIONS}

We have described in this paper is some detail an optical analogy of the twin barrier resonance experiment. The presentation has been made in three dimensions for an incoming gaussian laser beam characterized by a width of $\mbox{w}_{\0}=2\,\mbox{mm}$. The laser wavelength   has been set at $\lambda=633\,\mbox{nm}$ and the dielectric refractive index has been arbitrarily set at $n=\sqrt{3}$. The twin barriers are reproduced by  identical air intervals encompassing a narrow ($\sim$ mm) dielectric slab. The width sizes which allow for a clear angular separation of resonance peaks are of the order of several $\mu$m. Concentrating upon the first resonance peak, we find a very narrow angular spread comparable  to that of the laser beam itself. To clearly see the resonance peak, our primary objective in this proposal, we must have a laser spread of the order or even smaller than the resonance peak. In the contrary case, the laser beam would encompass multiple peaks and varying the incident angle $\alpha$ would not result in significant differences in the outgoing beam.

Once the resonance structure has been confirmed and assuming the numerical estimates of the previous sections are also convalidated, we can use the   angular separation between resonance peaks to measure any of the parameters assumed in this work.  For example the distance $\widetilde{D}$ or/and the barrier widths $\widetilde{L}$. The values of the resonance angles may also be used to measure the laser wavelength if all the other parameters are known. In principle, it should also be possible to detect variation in the twin nature of the barriers. This could be caused both by modifying the equal sizes of the barriers or by modifying one or both of the refractive indices of the air barriers, e.g. by density, temperature or chemical  nature of the ``air'' enclosed.   Also probable is the loss of resonance due to external disturbances such as vibrations. While these are annoying characteristics for any experiment they also open the possibility of practical applications.

\section*{\small \rm ACKNOWLEDGEMENTS}

 The authors  thank the anonymous
referees for their constructive comments and useful suggestions. One of the authors (SdL) also thanks the CNPq ({\em Bolsa de Produtividade em Pesquisa 2010/13}) and the FAPESP ({\em Grant No.\,10/02213-3}) for financial support.

\newpage

\begin{figure}[hbp]
\vspace*{5cm}
\begin{center}
    \setlength{\unitlength}{1mm}
    \begin{picture}(85,85)
        \put(-8,2){\line(1,0){56}}
        \put(-8,2){\line(0,1){80}}
        \put(48,2){\line(0,1){80}}
        \put(-8,82){\line(1,0){56}}

        \put(-9,1){\line(1,0){58}}
        \put(-9,1){\line(0,1){82}}
        \put(49,1){\line(0,1){82}}
        \put(-9,83){\line(1,0){58}}

        \put(2,70){\line(1,0){40}}
        \put(42,70){\line(0,-1){40}}
        \put(2,70){\line(1,-1){40}}

        \multiput(41,68)(-1,0){38}{$\cdot$}
        \multiput(41,67)(-1,0){1}{$\cdot$}
        \multiput(36,67)(-1,0){32}{$\cdot$}
        \multiput(41,66)(-1,0){1}{$\cdot$}
        \multiput(36,66)(-1,0){31}{$\cdot$}
        \multiput(41,65)(-1,0){1}{$\cdot$}
        \multiput(36,65)(-1,0){30}{$\cdot$}
        \multiput(41,64)(-1,0){34}{$\cdot$}
        \multiput(41,63)(-1,0){33}{$\cdot$}
        \multiput(41,62)(-1,0){32}{$\cdot$}
        \multiput(41,61)(-1,0){31}{$\cdot$}
        \multiput(41,60)(-1,0){30}{$\cdot$}
        \multiput(41,59)(-1,0){29}{$\cdot$}
        \multiput(41,58)(-1,0){28}{$\cdot$}
        \multiput(41,57)(-1,0){27}{$\cdot$}
        \multiput(41,56)(-1,0){26}{$\cdot$}
        \multiput(41,55)(-1,0){25}{$\cdot$}
        \multiput(41,54)(-1,0){24}{$\cdot$}
        \multiput(41,53)(-1,0){23}{$\cdot$}
        \multiput(41,52)(-1,0){22}{$\cdot$}
        \multiput(41,51)(-1,0){21}{$\cdot$}
        \multiput(41,50)(-1,0){20}{$\cdot$}
        \multiput(41,49)(-1,0){19}{$\cdot$}
        \multiput(41,48)(-1,0){18}{$\cdot$}
        \multiput(41,47)(-1,0){17}{$\cdot$}
        \multiput(41,46)(-1,0){16}{$\cdot$}
        \multiput(41,45)(-1,0){15}{$\cdot$}
        \multiput(41,44)(-1,0){14}{$\cdot$}
        \multiput(41,43)(-1,0){13}{$\cdot$}
        \multiput(41,42)(-1,0){12}{$\cdot$}
        \multiput(41,41)(-1,0){11}{$\cdot$}
        \multiput(41,40)(-1,0){10}{$\cdot$}
        \multiput(41,39)(-1,0){9}{$\cdot$}
        \multiput(41,38)(-1,0){8}{$\cdot$}
        \multiput(41,37)(-1,0){7}{$\cdot$}
        \multiput(41,36)(-1,0){6}{$\cdot$}
        \multiput(41,35)(-1,0){5}{$\cdot$}
        \multiput(41,34)(-1,0){4}{$\cdot$}
        \multiput(41,33)(-1,0){3}{$\cdot$}
        \multiput(41,32)(-1,0){2}{$\cdot$}
        \multiput(41,31)(-1,0){1}{$\cdot$}


        \put(2,60){\line(0,-1){40}}
        \put(42,20){\line(-1,0){40}}
        \put(2,60){\line(1,-1){40}}

       \multiput(2.5,20)(1,0){38}{$\cdot$}
        \multiput(2.5,21)(1,0){1}{$\cdot$}
        \multiput(7.5,21)(1,0){32}{$\cdot$}
        \multiput(2.5,22)(1,0){1}{$\cdot$}
        \multiput(7.5,22)(1,0){31}{$\cdot$}
        \multiput(2.5,23)(1,0){1}{$\cdot$}
        \multiput(7.5,23)(1,0){30}{$\cdot$}
        \multiput(2.5,24)(1,0){34}{$\cdot$}
        \multiput(2.5,25)(1,0){33}{$\cdot$}
        \multiput(2.5,26)(1,0){32}{$\cdot$}
        \multiput(2.5,27)(1,0){31}{$\cdot$}
        \multiput(2.5,28)(1,0){30}{$\cdot$}
        \multiput(2.5,29)(1,0){29}{$\cdot$}
        \multiput(2.5,30)(1,0){28}{$\cdot$}
        \multiput(2.5,31)(1,0){27}{$\cdot$}
        \multiput(2.5,32)(1,0){26}{$\cdot$}
        \multiput(2.5,33)(1,0){25}{$\cdot$}
        \multiput(2.5,34)(1,0){24}{$\cdot$}
        \multiput(2.5,35)(1,0){23}{$\cdot$}
        \multiput(2.5,36)(1,0){22}{$\cdot$}
        \multiput(2.5,37)(1,0){21}{$\cdot$}
        \multiput(2.5,38)(1,0){20}{$\cdot$}
        \multiput(2.5,39)(1,0){19}{$\cdot$}
        \multiput(2.5,40)(1,0){18}{$\cdot$}
        \multiput(2.5,41)(1,0){17}{$\cdot$}
        \multiput(2.5,42)(1,0){16}{$\cdot$}
        \multiput(2.5,43)(1,0){15}{$\cdot$}
        \multiput(2.5,44)(1,0){14}{$\cdot$}
        \multiput(2.5,45)(1,0){13}{$\cdot$}
        \multiput(2.5,46)(1,0){12}{$\cdot$}
        \multiput(2.5,47)(1,0){11}{$\cdot$}
        \multiput(2.5,48)(1,0){10}{$\cdot$}
        \multiput(2.5,49)(1,0){9}{$\cdot$}
        \multiput(2.5,50)(1,0){8}{$\cdot$}
        \multiput(2.5,51)(1,0){7}{$\cdot$}
        \multiput(2.5,52)(1,0){6}{$\cdot$}
        \multiput(2.5,53)(1,0){5}{$\cdot$}
        \multiput(2.5,54)(1,0){4}{$\cdot$}
        \multiput(2.5,55)(1,0){3}{$\cdot$}
        \multiput(2.5,56)(1,0){2}{$\cdot$}
        \multiput(2.5,57)(1,0){1}{$\cdot$}

        \put(2,67.5){\line(1,-1){40}}
        \put(2,62.5){\line(1,-1){40}}
        \put(2,62.5){\line(0,1){5}}
        \put(42,22.5){\line(0,1){5}}

         \multiput(2,62.5)(1,-1){19}{$\cdot$}
         \multiput(2,63.5)(1,-1){19}{$\cdot$}
         \multiput(2,64.5)(1,-1){20}{$\cdot$}
         \multiput(2,65.5)(1,-1){21}{$\cdot$}

          \multiput(23,41.5)(1,-1){19}{$\cdot$}
          \multiput(24,41.5)(1,-1){18}{$\cdot$}
          \multiput(25,41.5)(1,-1){17}{$\cdot$}
          \multiput(25,42.5)(1,-1){17}{$\cdot$}

 \thicklines

       \put(0,75){\line(2,-1){10}} \put(0,75){\vector(2,-1){5}}
       \put(10,70){\line(1,-2){8}}  \put(10,70){\vector(1,-2){4}}
       \put(10,70){\line(2,1){20}}   \put(10,70){\vector(2,1){10}}

           \put(17.5,53){$\bullet$}
           \put(19.5,41){$\bullet$}

         \put(20,42){\line(1,-2){11}} \put(20,42){\vector(1,-2){6}}
         \put(31,20){\line(2,-1){12}} \put(31,20){\vector(2,-1){7}}
         \put(31,20){\line(1,2){2.5}} \put(31,20){\vector(1,2){1.75}}

          \put(38,66){$\boldsymbol{n}$}
           \put(21.8,43.6){$\boldsymbol{n}$}
           \put(4.5,22){$\boldsymbol{n}$}

            \put(60,5){ \fbox{\bf Fig.\,1}}

           \put(39,76){\bf Air}

         \put(5,14){\bf Resonant }
          \put(5,10){\bf Laser}
           \put(5,6){\bf Tunneling}

 \multiput(-3,68)(0,-1){58}{$\cdot$}
\put(-2.6,12){\vector(0,-1){2}}
          \put(-3.6,7.5){$\boldsymbol{z_*}$}

 \put(-7,69){$a_*$}  \put(-4,69){$--$}

 \put(-7,19){$b_*$}  \put(-4,19){$--$}

 \multiput(-6,72)(1,2){4}{$\cdot$}
 \put(-2.2,79.6){\vector(1,2){1}}
  \put(-0.6,79.4){$\boldsymbol{y}$}

\put(3.9,74.5){$\boldsymbol{z}$}

\multiput(35,71)(1,0){4}{$\cdot$}
 \put(40.5,72){\vector(1,0){1}}
  \put(42,71.5){$\boldsymbol{y_*}$}

\end{picture}
\begin{picture}(30,70)
        \put(-2,0){\line(1,0){30}}
        \put(-2,0){\line(0,1){80}}
        \put(28,80){\line(0,-1){80}}
        \put(-2,72){\line(1,0){30}}
         \put(-2,80){\line(1,0){30}}

        \put(2,60){\line(1,-1){22}}
        \put(2,52){\line(1,-1){22}}
        \put(2,35){\line(1,-1){22}}
        \put(2,27){\line(1,-1){22}}

        \put(1,41){{\large $\boldsymbol{n}$}}
        \put(1,63){{\large $\boldsymbol{n}$}}
        \put(1,19.8){{\large $\boldsymbol{n}$}}

        \put(0,54.5){ $\boldsymbol{1}$}
        \put(1,29.5){$\boldsymbol{1}$}

       \put(10,44.25){ $\boldsymbol{\widetilde{L}}$}
       \put(9.8,32.5){$\boldsymbol{\widetilde{D}}$}
       \put(10,20){$\boldsymbol{\widetilde{L}}$}

   \multiput(2,60.2)(1,-1){22}{$\cdot$}
   \multiput(3,61)(1,-1){21}{$\cdot$}
   \multiput(4,62)(1,-1){20}{$\cdot$}
   \multiput(2,66)(1,-1){22}{$\cdot$}
   \multiput(2,68)(1,-1){22}{$\cdot$}
    \multiput(4,68)(1,-1){20}{$\cdot$}
     \multiput(6,68)(1,-1){18}{$\cdot$}
      \multiput(8,68)(1,-1){16}{$\cdot$}
       \multiput(10,68)(1,-1){14}{$\cdot$}
    \multiput(12,68)(1,-1){12}{$\cdot$}
     \multiput(14,68)(1,-1){10}{$\cdot$}
      \multiput(16,68)(1,-1){8}{$\cdot$}
       \multiput(18,68)(1,-1){6}{$\cdot$}
    \multiput(20,68)(1,-1){4}{$\cdot$}
     \multiput(22,68)(1,-1){2}{$\cdot$}

       \multiput(2,50)(1,-1){22}{$\cdot$}
     \multiput(2,48)(1,-1){22}{$\cdot$}
      \multiput(2,46)(1,-1){22}{$\cdot$}
       \multiput(2,44)(1,-1){9}{$\cdot$}
    \multiput(4,40)(1,-1){6}{$\cdot$}
     \multiput(3,39)(1,-1){6}{$\cdot$}
    \multiput(14,32)(1,-1){10}{$\cdot$}
     \multiput(13,31)(1,-1){11}{$\cdot$}
      \multiput(12,30)(1,-1){12}{$\cdot$}
    \multiput(2,38)(1,-1){22}{$\cdot$}
     \multiput(2,36)(1,-1){22}{$\cdot$}

        \multiput(2,25)(1,-1){22}{$\cdot$}
         \multiput(2,23)(1,-1){22}{$\cdot$}
          \multiput(4,19)(1,-1){18}{$\cdot$}
           \multiput(3,18)(1,-1){17}{$\cdot$}
            \multiput(2,17)(1,-1){16}{$\cdot$}
             \multiput(2,15)(1,-1){14}{$\cdot$}
              \multiput(2,13)(1,-1){12}{$\cdot$}
               \multiput(2,11)(1,-1){10}{$\cdot$}
                \multiput(2,9)(1,-1){8}{$\cdot$}
                \multiput(2,7)(1,-1){6}{$\cdot$}
             \multiput(2,5)(1,-1){4}{$\cdot$}
              \multiput(2,3)(1,-1){2}{$\cdot$}

\thicklines
\put(-10,35){\vector(0,-1){14.14}}
      \put(-10,35){\vector(-1,-1){10}}

\put(-11,17.5){$\boldsymbol{z_*}$}
\put(-22,21){$\boldsymbol{\widetilde{z}}$}

\put(1,75 ){\bf Amplification}

\thicklines
      \put(11.8,31){\vector(1,1){5.5}}
      \put(13.8,33){\vector(-1,-1){5}}

      \put(16.75,45.3){\vector(-1,-1){4}}
      \put(12.8,41.3){\vector(1,1){4}}

      \put(15.75,21.3){\vector(-1,-1){4}}
      \put(11.8,17.3){\vector(1,1){4}}
\end{picture}
\end{center}

 \caption{Schematic presentation of a double prism
plus a narrow slab separated by air gaps.}
\end{figure}
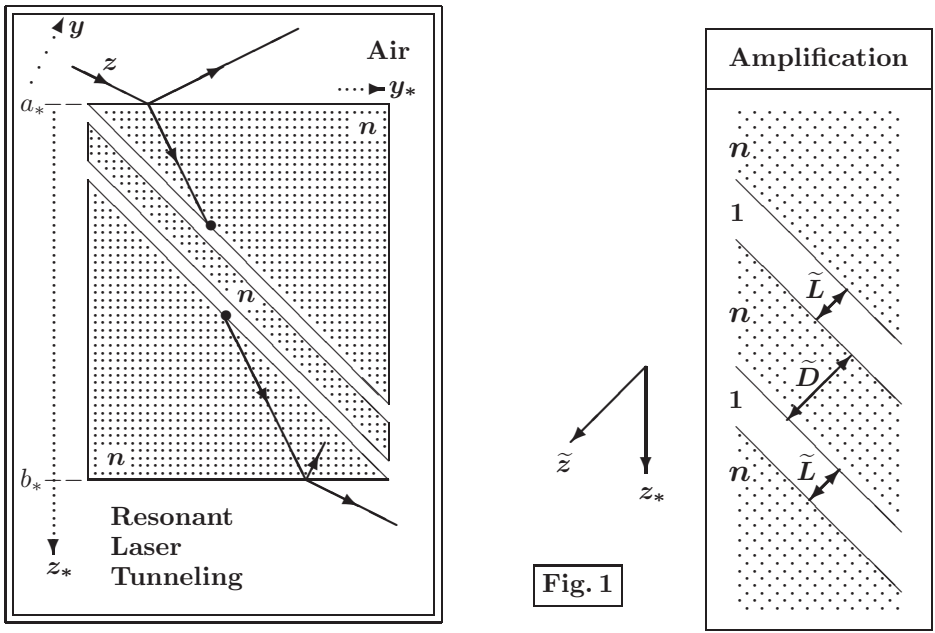

\begin{figure}[hbp]
\vspace*{-2cm}
\begin{center}
    \setlength{\unitlength}{1mm}
    \begin{picture}(100,80)
        \put(-2,0){\line(1,0){100}}
        \put(-2,0){\line(0,1){82}}
        \put(-2,82){\line(1,0){100}}
        \put(48,0){\line(0,1){82}}
        \put(98,0){\line(0,1){82}}

        \put(-3,-1){\line(1,0){102}}
        \put(-3,-1){\line(0,1){84}}
        \put(-3,83){\line(1,0){102}}
        \put(99,-1){\line(0,1){84}}
        \thicklines
        \put(2,70){\line(1,0){40}}
        \put(42,70){\line(0,-1){40}}
        \put(2,70){\line(1,-1){40}}

        \put(52,70){\line(1,0){40}}
        \put(52,70){\line(0,-1){40}}
        \put(52,30){\line(1,1){40}}


       \put(0,75){\line(2,-1){10}} \put(0,75){\vector(2,-1){5}}
       \put(10,70){\line(1,-2){8}}  \put(10,70){\vector(1,-2){4}}
       \put(10,70){\line(2,1){20}}   \put(10,70){\vector(2,1){10}}
       \put(18,54){\line(2,-1){14}} \put(18,54){\vector(2,-1){8}}

        \thinlines

        \put(7.5,43.5){\vector(-1,-1){1}}
         \multiput(18.5,54)(-1,-1){12}{$\cdot$}
         \put(17.2,53){$\bullet$}
         \put(5,39.5){$\boldsymbol{\tilde{z}}$}
          \multiput(9.5,62)(0,-1){6}{$\cdot$}
          \put(9.9,57){\vector(0,-1){1}}
          \put(8.8,53.5){$\boldsymbol{z_*}$}
          \put(38,66){$\boldsymbol{n}$}

           \put(39,76){\bf Air}

           \thicklines
       \put(50,75){\line(2,-1){10}} \put(50,75){\vector(2,-1){5}}
       \put(60,70){\line(1,-2){10.6}}  \put(60,70){\vector(1,-2){5.3}}
        \put(70,47.7){$\bullet$}
        \put(71,48.3){\line(-2,1){13}}  \put(71,48.3){\vector(-2,1){6.5}}
       \put(60,70){\line(2,1){20}}   \put(60,70){\vector(2,1){10}}


          \put(54,66){$\boldsymbol{n}$}

           \put(89,76){\bf Air}


         \multiput(9.5,70)(0,1){7}{$\cdot$}
         \put(6.3,72.5){$\boldsymbol{\alpha}$}
         \multiput(9.5,70)(0,-1){9}{$\cdot$}
         \put(12.5,66.5){$\boldsymbol{\frac{\pi}{2}-\beta}$}
         \put(38.95,34){$\boldsymbol{\frac{\pi}{4}}$}
         \multiput(18.5,54)(1,1){7}{$\cdot$}
          \put(17.4,56.5){$\boldsymbol{\gamma}$}

         \put(2,28){$\boldsymbol{n > 1}$}
          \put(3,22){$\boldsymbol{\sin \gamma_{\c} = 1\,/\,n}$}
          \put(11.5,10){\shadowbox{$\boldsymbol{\gamma = \frac{\pi}{4}+ \beta}$}}
          \put(2,4){\bf  TIR\,:\,\,\, $\boldsymbol{\gamma > \gamma_{\c}}$}
           \put(42,2){\bf  (a)}

         \multiput(59.5,70)(0,1){7}{$\cdot$}
         \put(56.3,72.5){$\boldsymbol{\alpha}$}
         \multiput(59.5,70)(0,-1){9}{$\cdot$}
         \put(62.5,66.5){$\boldsymbol{\frac{\pi}{2}-\beta}$}
         \put(67.5,55.8){$\boldsymbol{\frac{\pi}{2}-\gamma}$}
         \put(52.3,34){$\boldsymbol{\frac{\pi}{4}}$}
          \multiput(69.6,48.3)(-1,1){8}{$\cdot$}

          \put(61.5,10){\shadowbox{$\boldsymbol{\gamma= \frac{\pi}{4}- \beta}
          $}}
           \put(49.2,2){\bf  (b)}
 \put(84,3){ \fbox{\bf Fig.\,2}}
\end{picture}
\end{center}

 \caption{Schematic presentation of a diagonally
sliced dielectric prism. In the case (a), the condition $n>\sqrt{2}$ guarantees TIR for all the incidence angles ($0\leq \sin \alpha \leq 1$). In the case (b), we have TIR for $n>\sqrt{2}$ and
$0\leq \sin \alpha \leq \sqrt{n^{^2}-2\,\sqrt{n^{^2}-1}}/\sqrt{2}$.}
\end{figure}
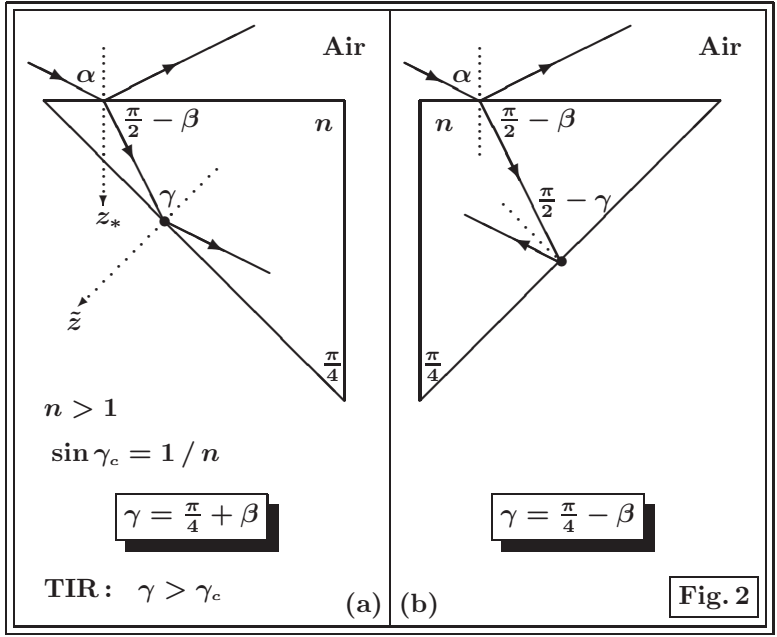

\newpage

\begin{figure}[hbp]
\hspace*{-1.35cm}
\includegraphics[width=16.5cm, height=20cm, angle=0]{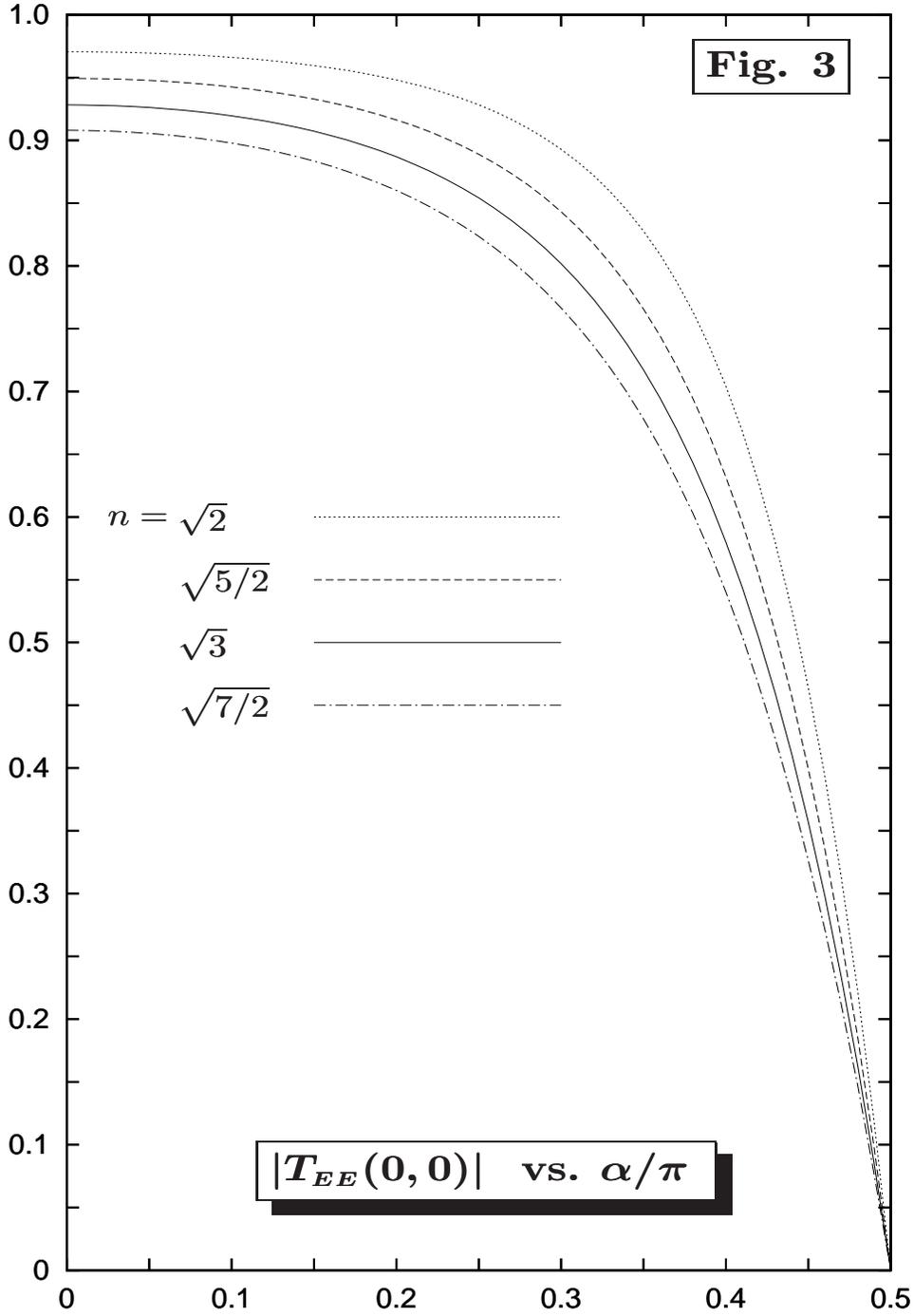}
\vspace*{-1.2cm}
 \caption{The modulus of the entry exit transmission amplitude for the  central (s-polarized) laser ray, $k_x=k_y=0$, plotted as a function of the incident angle, $\alpha$, for different values of refractive index, $n$. The continuous line corresponds to the refractive index, $n=\sqrt{3}$,  used in our numerical studies. For small incidence angles, the variation is very smooth and less than $10\%$.}
\end{figure}

\newpage

\begin{figure}[hbp]
\hspace*{-1.35cm}
\includegraphics[width=16.5cm, height=20cm, angle=0]{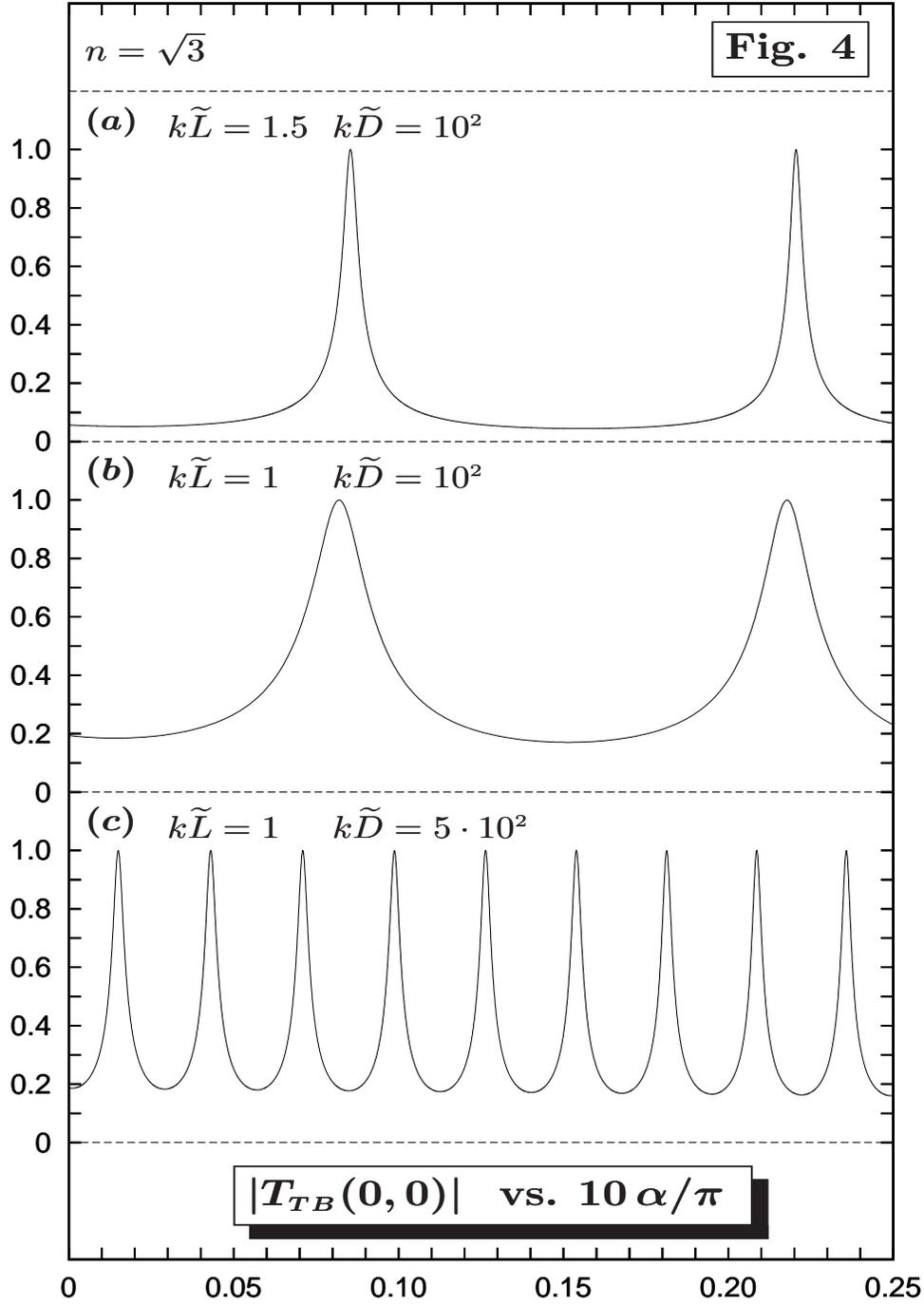}
\vspace*{-1.2cm}
 \caption{The modulus of the twin barrier transmission amplitude for the central (s-poralized) laser ray, $k_x=k_y=0$, plotted as a function of the incident angle, $\alpha$, for different values of the twin air gaps, $\widetilde{L}$, and the dielectric slab, $\widetilde{D}$. The resonance peaks are narrower as the air gaps increase and the distance between any two peaks decreases by increasing the dielectric slab dimension.}
\end{figure}

\newpage

\begin{figure}[hbp]
\hspace*{-1.35cm}
\includegraphics[width=16.5cm, height=22cm, angle=0]{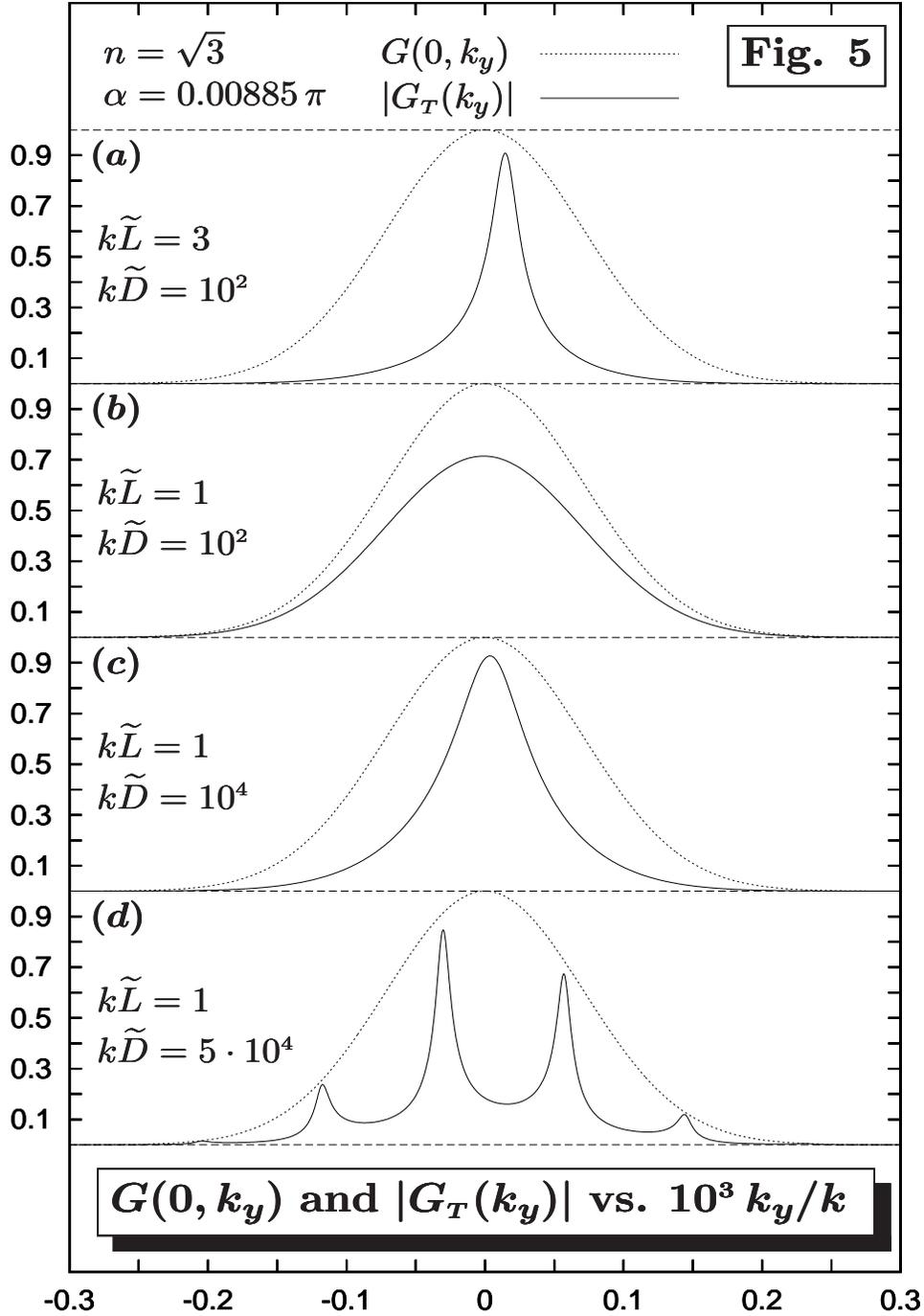}
\vspace*{-1.2cm}
 \caption{The incoming gaussian distribution, $G(0,k_y)$,  and the transmitted distribution, $|G_{\T}(k_y)|$, for s-poralized laser beams of size $\mbox{w}_{\0}=2\,\mbox{mm}$ are plotted as a function of $k_y$ for different values of the twin air gaps, $\widetilde{L}$, and the dielectric slab, $\widetilde{D}$. The incident angle, $\alpha$, is chosen around the first resonance for the first three cases, see plots (a,b,c). In the last plot (d), it is manifest the presence of multiple peaks.}
\end{figure}

\newpage

\begin{figure}[hbp]
\hspace*{-1.35cm}
\includegraphics[width=16.5cm, height=22cm, angle=0]{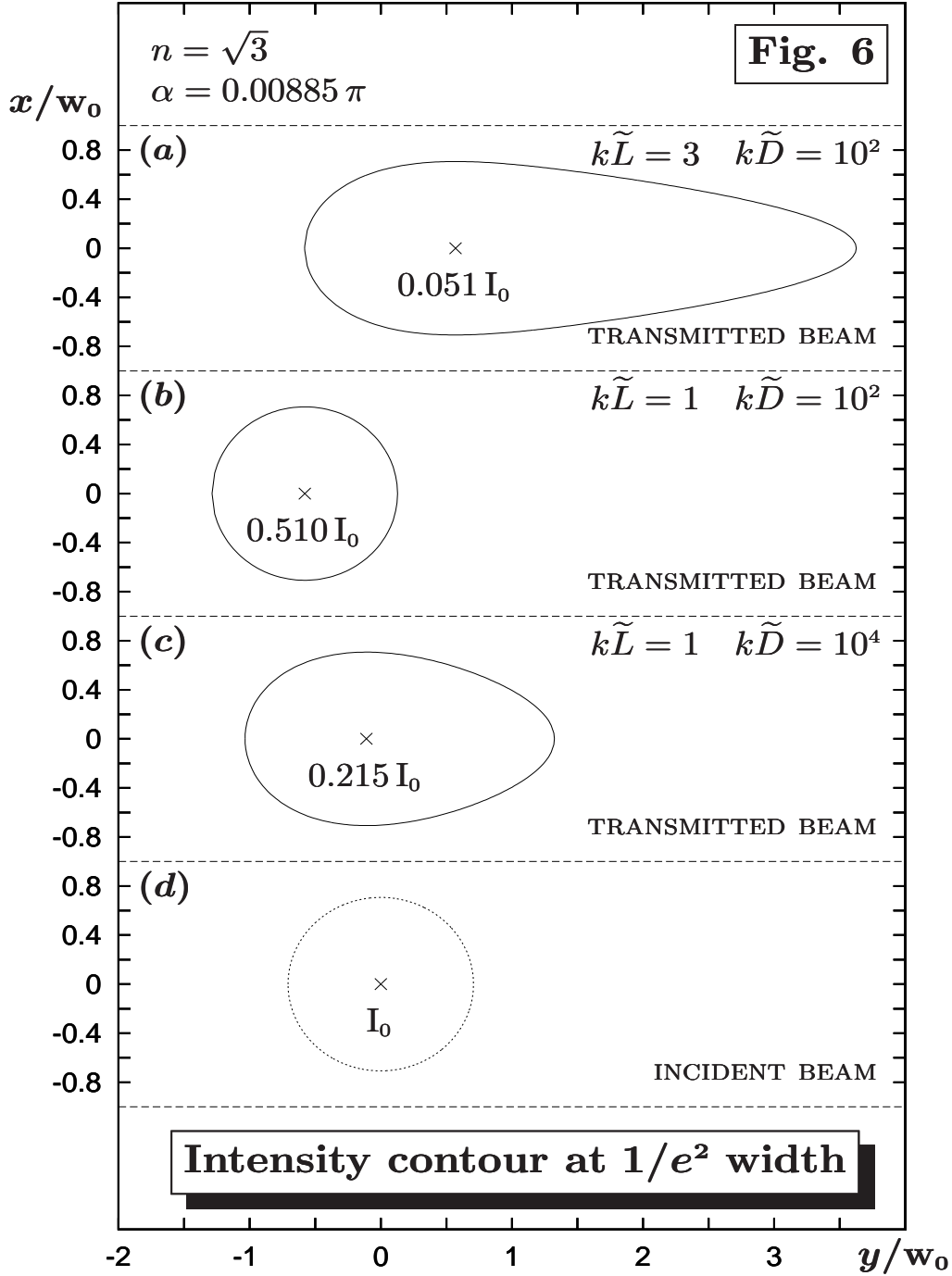}
\vspace*{-1.2cm}
 \caption{The exit contour of the (s-poralized) laser intensity at $1/e^{\2}$ width is plotted in the $x$-$y$ plane for different values of the twin air gaps, $\widetilde{L}$, and the dielectric slab, $\widetilde{D}$. The distance from the source of the laser, $z$-axis,  and the dielectric block dimension, $z_*$-axis, are respectively chosen to be $500$ w$_{\0}$ ($1$ m) and  $50$ w$_{\0}$ ($10$ cm). The values of the peak intensities are given in terms of the incoming laser beam peak intensity I$_{\0}$.}
\end{figure}

\end{document}